\documentclass[aps,prl,floats,twocolumn,floatfix,showpacs]{revtex4}
\usepackage{epsf,graphicx}
\usepackage{amssymb}
\usepackage{amsmath}
\usepackage{eepic}
\newlength{\piclen}
\begin{document}

\title{Universal Behavior in Heavy Electron Materials}

\author{Yi-feng Yang and David Pines}
\affiliation{Los Alamos National Laboratory, Los Alamos, NM 87545 and\\
Department of Physics, University of California, Davis, CA 95616}
 \date{November 02, 2007}

\begin{abstract}
We present our finding that an especially simple scaling 
expression describes the formation of a new state of quantum matter, 
the Kondo Fermi liquid (KL) in heavy electron materials. Emerging at 
$T^*$ as a result of the collective coherent hybridization of localized 
f electrons and conduction electrons, the KL possesses a non-Landau 
density of states varying as $(1-T/T^*)^{3/2}[1+\ln(T^*/T)]$. We show 
that four independent experimental probes verify this scaling behavior 
and that for CeIrIn$_5$ the KL state density is in excellent agreement 
with the recent microscopic calculations of hybridization in this material 
by Shim, Haule, and Kotliar. 
\end{abstract}

% 71.27.+a  Strongly correlated electron systems; heavy fermions
% 71.10.Fd  Lattice fermion models (Hubbard model, etc.)
% 71.30.+h  Metal-insulator transitions and other electronic transitions
% 71.20.Eh  Rare earth metals and alloys
% 75.20.Hr  Local moment in compounds and alloys; Kondo effect, valence
%           fluctuations, heavy fermions
% 75.47.Gk  Colossal magnetoresistance

\maketitle

Scaling behavior in heavy electron materials was first suggested 
by Nakatsuji {\em et al.}~\cite{Nakatsuji2004}, 
hereafter NPF, who found, through a careful 
analysis of experiments \cite{Nakatsuji2002} 
on the thermal and magnetic behavior of 
the 115 materials, Ce$_{1-x}$La$_x$CoIn$_5$ and its Ir counterpart, 
that these results could be interpreted using a two-fluid model of 
their specific heat and bulk susceptibility. One fluid, the emergent 
itinerant heavy electron component, was characterized by an order 
parameter, $f(T/T^*)$, that increased with decreasing temperature and
scaled with the temperature, $T^*$, at which itinerant behavior emerged;
the second component was a spin liquid, characterized by an order 
parameter, $[1-f(T/T^*)]$, and made up of weakly interacting local moments, that 
at low temperatures could be described by a collection of non-interacting 
Kondo impurities. The itinerant component possessed two striking properties: 
its spin susceptibility and specific heat were related by a temperature 
independent Wilson ratio, $R_W$, and both exhibited scaling behavior, 
varying logarithmically as $T^*/T$. The NPF proposal was subsequently 
shown to extend to a broad spectrum of heavy electron materials by 
Curro {\em et al.} \cite{Curro2004}, hereafter CYSP, who showed that 
the NPF two-fluid model provided a natural explanation for the hyperfine 
anomaly found in materials for which NMR and $\mu$SR measurements of 
the Knight shift (KS) do not track the measured bulk spin susceptibility. 
By interpreting its appearance as the onset of two-fluid behavior, their 
work provided an independent determination of $T^*$ and of the heavy 
electron spin susceptibility that emerges below $T^*$, although for 
CeCoIn$_5$, the NMR expression for this quantity differed somewhat from 
its NPF thermal determination. Taken together, 
the work of these authors suggests that in a Kondo lattice the 
hybridization of the localized f-electron magnetic moments with the 
conduction electrons becomes a collective or global process at $T^*$ that 
leads to the formation of a new state of quantum matter, the Kondo Fermi 
liquid, or KL, whose density of states displays distinctly non-Landau 
scaling behavior between $T^*$ and a typically much lower cut-off 
temperature, $T_0$, below which another form of quantum order appears.
As we shall see, it co-exists there with both the spin liquid and 
a "light" electron Landau Fermi liquid made of those conduction 
electrons that do not hybridize with the local moments.

In this letter, we present the results of an analysis that unites 
the thermal and magnetic determinations of the parameters that characterize 
KL behavior and leads to an especially simple scaling description 
of its density of states. We show that this KL density of states 
is seen in a number of other experimental probes of heavy electron behavior, 
and that quite remarkably, when our scaling expression for the KL state density 
is applied to the heavy electron material CeIrIn$_5$, it yields a state density 
that is in excellent agreement with the recent microscopic calculations for this 
material by Shim {\em et al.}~\cite{Shim2007}. The agreement provides a double validation
--- of the methods used in their microscopic approach and the results obtained 
here using a phenomenological approach --- and we discuss its implications for 
heavy electron behavior above $T_0$, and the new quantum ordered states that 
emerge below $T_0$.

Our approach to reconciling small differences in the details of the scaling 
results of NPF and CYSP for the Co 115 material, and obtaining a simple form for 
its KL density of states, is perhaps best explained by comparing the NPF 
two fluid description of the bulk susceptibility:
\begin{equation}
\chi=f(T)\chi_{KL}+[1-f(T)]\chi_{SL},
\label{Eq:NPFchi}
\end{equation}
in which a Fermi liquid contribution from those "light" conduction electrons 
that do not hybridize is neglected, and the comparable expression for the Knight 
shift of a probe nucleus (apart from a constant offset $K_0$),
\begin{equation}
K-K_0=A f(T)\chi_{KL} + B[1-f(T)]\chi_{SL},
\label{Eq:CurroK}
\end{equation}
in which the spin liquid component is coupled to the probe nucleus by a 
transferred hyperfine coupling constant $B$, while the itinerant Kondo liquid 
component is coupled by a direct hyperfine coupling $A$. The emergent anomalous 
component of the Knight shift considered in CYSP can then be written as
\begin{equation}
K_{anom}=K-K_0-B\chi=(A-B)f(T)\chi_{KL}.
\label{Eq:Kanom}
\end{equation}
On combining Eqs.~(\ref{Eq:NPFchi}) and (\ref{Eq:Kanom}), one gets a very simple 
formula for the order parameter $f(T)$,
\begin{equation}
f(T)=1-\frac{\chi(T)-\alpha K_{anom}(T)}{\chi_{SL}(T)},
\label{Eq:fT}
\end{equation}
in which $\alpha=1/(A-B)$ and the only unknown quantity is $\chi_{SL}$. The behavior 
of $\chi_{SL}$ is known in two limits: at high temperatures, $T>T^*$, $\chi_{SL}$ is given
by the Curie-Weiss law, $\chi_{CW}(T)\!\sim\!(T+T^*)^{-1}$; at low temperatures, $T<T^*/4$, 
say, because the local f-moments are strongly screened by the weakly hybridized conduction 
electrons, it reduces to a collection of non-interacting single Kondo impurities, an 
assumption born out by the NPF results. 
\begin{figure}[t]
{\includegraphics[width=8.6cm,angle=0]{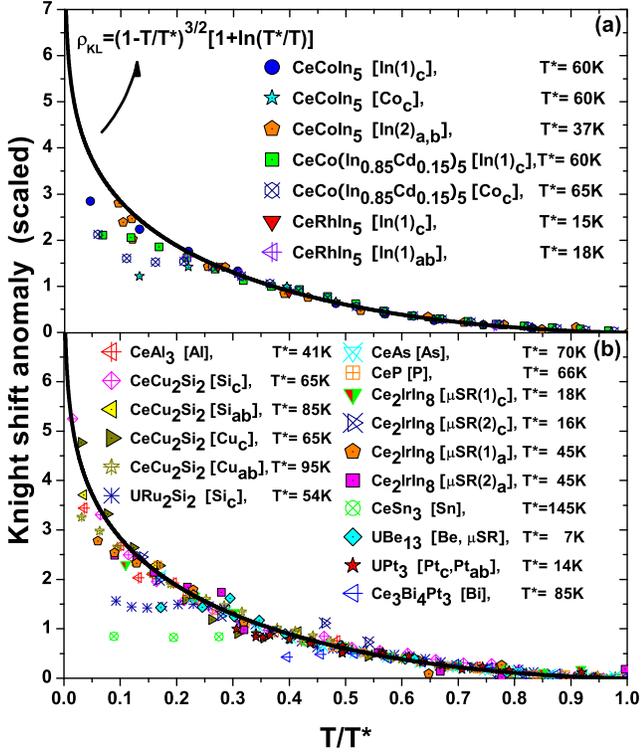}}
\caption{(Color online)
{Scaling behavior of the anomalous Knight shift for Ce-115 and some other heavy electron 
materials (data from Refs.~\cite{Heffner2005,Heffner2007} and references in Ref.~\cite{Curro2004}). 
The solid lines represent the KL density of states.}
\label{Fig:KS}}
\end{figure}
A simple linear interpolation between these limits then yields $\chi_{SL}$, while 
$\alpha$ is determined by comparing the results obtained with above
approach with those found by a self-consistent determination of f(T) based on the NPF 
approach. The result we obtain for the order parameter,
\begin{equation}
f(T/T^*)=f(0)\left(1-\frac{T}{T^*}\right)^{3/2},
\label{Eq:KLfT}
\end{equation}
can be combined with our result for the KL quasiparticle 
effective mass determined from the specific heat and Knight shift fits,
\begin{equation}
m^*_{KL}=m_h \left(1+\ln\frac{T^*}{T}\right),
\label{Eq:KLmass}
\end{equation}
seen in $\chi_{KL}$, where $m_h$ is the quasiparticle effective mass at $T^*$, to yield 
the reduced density of states for the Kondo liquid,
\begin{equation}
\rho_{KL} = \left(1-\frac{T}{T^*}\right)^{3/2}\left(1+\ln\frac{T^*}{T}\right).
\label{Eq:KLdos}
\end{equation}
Fig.~\ref{Fig:KS} shows that our result provides an excellent fit to the values of 
$K_{anom}$ obtained from NMR experiments for both the 115 family of heavy electron materials
and all others that display a Knight shift anomaly.

Three recent experiments on the 115 materials provide additional 
evidence for the presence of the Kondo liquid and a description of its density of states 
using Eq.~(\ref{Eq:KLdos}): the anomalous Hall effect, tunneling experiment, and 
Raman scattering.

{\bf\it Anomalous Hall effect}

\begin{figure}[t]
{\includegraphics[width=8.6cm,angle=0]{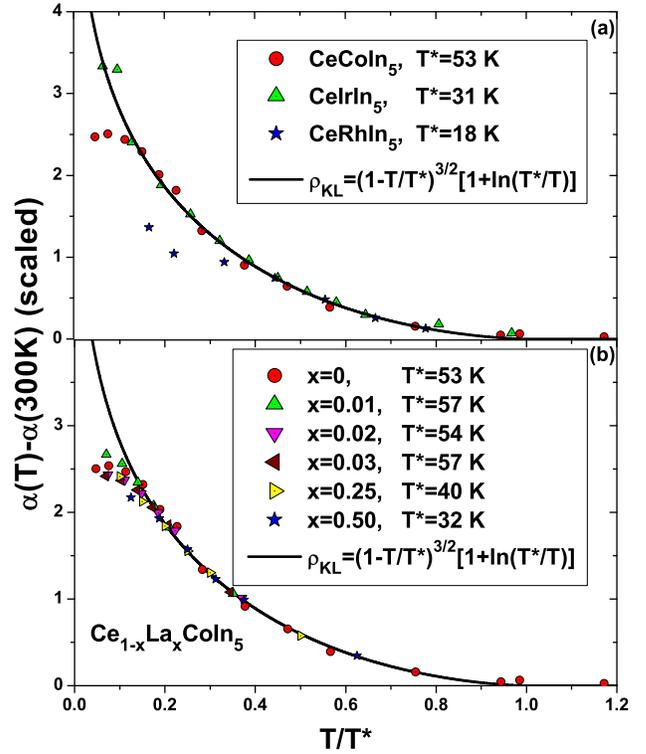}}
\caption{(Color online)
{Scaling behavior of the anomalous Hall coefficient for (a) Ce-115 materials and 
(b) Ce$_{1-x}$La$_x$CoIn$_5$. The solid lines are the KL density of states.}
\label{Fig:Hall}}
\end{figure}
Hundley {\em et al.} \cite{Hundley2004} have found that the Hall effect in the Ce-115 
compounds does not follow the standard skew scattering theory, based on single Kondo 
impurity scattering, that explains the behavior of many heavy electron materials 
\cite{Fert1987}. In this theory, the influence of applied external fields on the local 
Kondo impurities produces a left-right asymmetry that is seen as a skew scattering 
contribution to the Hall effect. To understand the departure of their results from this 
theory, Hundley {\em et al.} introduce the dimensionless function,
\begin{equation}
\alpha(T)=\left( R^{Ce}_H(T) - R^{skew}_H(T) \right)/R^{La}_H(T),
\label{Eq:Hall}
\end{equation}
to describe additional effects originated from the localized Ce ions. Here 
$R^{Ce}_H(T)$ is the Hall coefficient of the Ce-115 compound and $R^{La}_H(T)$ 
is the Hall coefficient of the corresponding La compound. Hundley {\em et al.}~find 
that the skew scattering term 
$R^{skew}_H(T)$ makes a minor contribution to the total Hall effect and that while 
$\alpha(T)$ is nearly constant at high temperatures, it increases sharply below 
a characteristic onset temperature that we identify with T$^*$. On analysing 
their results, we find that $\alpha(T)$ exhibits exactly the following scaling 
behavior:
\begin{equation}
\alpha(T)-\alpha(300K) \sim \rho_{KL}(T/T^*).
\label{Eq:alpha}
\end{equation}
\begin{figure}[t]
{\includegraphics[width=7.5cm,angle=0]{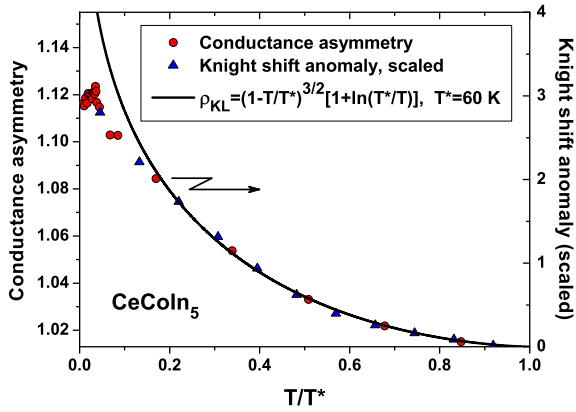}}
\caption{(Color online)
{Scaling behavior of the conductance asymmetry in the tunneling experiment
on CeCoIn$_5$. The KS anomaly is also shown for comparison. The solid line 
indicates the KL density of states.}
\label{Fig:Andreev}}
\end{figure}
As may be seen in Fig.~\ref{Fig:Hall}(a), the scaled $\alpha(T)-\alpha(300K)$ for all 
three Ce-115 materials falls upon the universal density of states of the Kondo liquid 
when one takes the onset temperature $T^*$ for the anomalous Hall effect to be 53\,K 
for CeCoIn$_5$, 31\,K for CeIrIn$_5$, and 18\,K for CeRhIn$_5$, results that are 
comparable to the values of $T^*$ for these materials obtained from an analysis 
of their magnetic and thermal behavior. As shown in Fig.~\ref{Fig:Hall}(b), 
similar scaling behavior is also found in Ce$_{1-x}$La$_x$CoIn$_5$ \cite{Hundleyunpub}. 
In these materials, the onset temperature of the Hall anomaly again agrees very 
well with the earlier results from the magnetic susceptibility and the specific heat 
and displays clearly the linear decrease of $T^*$ with increasing La doping that was 
found by NPF. We conclude that Hall measurements not only provide a direct measure 
of the KL state density, but offer a reliable independent estimate of $T^*$.

{\bf\it Tunneling experiment}

Kondo liquid behavior has also been found to play a role in tunneling conductance. In 
the recent experiment by Park {\em et al.}~on CeCoIn$_5$ \cite{Park2007}, a conductance 
asymmetry, defined as the ratio of the differential conductance at -2 mV and 2 mV, is 
found to develop at the characteristic KL temperature $T^*$ and to increase 
with decreasing temperature down to the superconducting transition temperature 
$T_c$ (2.3 K). As Curro \cite{Curroprivate} has pointed out, their measured asymmetry 
follows the KS anomaly measured by CYSP and as may be seen in Fig.~\ref{Fig:Andreev}, 
both follow the universal KL density of states for temperatures above the cut-off 
temperature, $T_0 \approx 10\,$K, that marks the end of Kondo scaling behavior, 
while their mutual agreement continues down to even lower temperature. This tells us 
that the breakdown of universality does not necessarily indicate the breakdown of the two 
fluid description, but signals instead a transition of the KL state to some other low 
temperature quantum state. 

{\bf\it Raman scattering}

\begin{figure}[t]
{\includegraphics[width=7.5cm,angle=0]{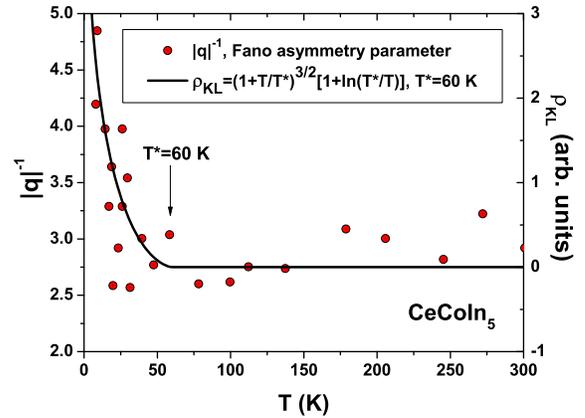}}
\caption{(Color online)
{Fano asymmetry parameter $|q|^{-1}$ from the Raman spectra of CeCoIn$_5$ compared to 
the KL density of states.}
\label{Fig:Raman}}
\end{figure}

Very recently, Martinho {\em et al.}~\cite{Martinho2007} have used Raman scattering 
to probe the behavior of CeCoIn5. They note that the Fano line shape of the Raman 
spectral function probes the coupling of phonons to itinerant electrons and that 
the Fano asymmetry parameter q becomes markedly temperature dependent below a 
temperature of order $T^*$. This led us to compare it with our KL state density, 
with the results shown in Fig.~\ref{Fig:Raman}. While not as convincing as the 
anomalous Hall effect or tunneling conductance asymmetry, these results suggest 
that in the future Raman scattering might offer a useful probe of Kondo liquid behavior. 

\begin{figure}[t]
{\includegraphics[width=7.7cm,angle=0]{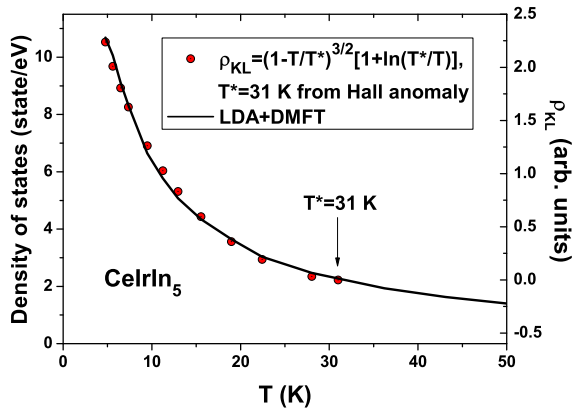}}
\caption{(Color online)
{LDA+DMFT density of states of the quasiparticles in CeIrIn$_5$ compared to the 
KL density of states. We take the onset temperature $T^*=31\,$K from the Hall anomaly.}
\label{Fig:DOS}}
\end{figure}

At this stage it is natural to inquire what, if any, theoretical basis
exists for our phenomenological description of Kondo liquid behavior.
The answer comes in the very recent work of Shim {\em et al.} \cite{Shim2007},
hereafter SHK, who carry out a first principles calculation of the localized to 
itinerant transition in CeIrIn$_5$ using Dynamical Mean Field Theory
in combination with Local Density Approximation (LDA+DMFT). 
In Fig.~\ref{Fig:DOS} we compare their calculated density of states at the 
quasiparticle peak with our universal KL density of states; $T^*$ is chosen to be 31 K, as 
suggested by the anomalous Hall data. Since in the SHK result there is no clear onset 
temperature at which the density of states starts to increase, we simply compare the 
low temperature ($T<T^*$) portions and find a linear relation between them.

The excellent agreement between the two results indicates
that SHK have captured the Kondo liquid state density without
having to resort to a cluster calculation below $T^*$, a truly remarkable
finding that provides quite strong evidence for the validity of both their
microscopic approach to hybridization and our phenomenological scaling 
expression for the KL state density. Since there is no experimental evidence 
for a change in the specific heat at the hyrbridization phase transition
(i.e.~nothing special happens at $T^*$), and since local hybridization can
influence the density of states up to very high temperatures, it is to be
expected that an exact calculation of the state density would do just what
their calculation does, i.e.~go smoothly through $T^*$ and yield our scaling
results below $T^*$. 

It is useful to make a distinction between the local (i.e.~single Kondo impurity) 
hybridization that must begin at temperatures well above $T^*$ and the 
universal global hybridization behavior that begins at $T^*$. The local 
hybridization is expected to be highly anisotropic and complex because the 
Kondo coupling, $J$, of the single Kondo impurities to different parts of the 
Fermi surface will be different; this is just what SHK find.  $T^*$ is the 
temperature at which, as a result of feedback effects on the coupling between 
individual hybridization channels, hybridization becomes a "simple"
collective coherent universal phenomenon, characterized by an order parameter, 
$f(T/T^*)$, that goes as $(1-T/T^*)^{3/2}$, and a logarithmic increase in the average 
effective mass, $m^*_{KL}=m_h[1 + \ln(T^*/T)]$, where $m_h$ incorporates the 
bybridization occurring above $T^*$. We further note that NPF argue that the 
physical origin of the energy scale $T^*$ that characterizes KL behavior is 
the nearest neighbor coupling between the localized electrons, a view that is 
supported by its appearance as the Curie-Weiss parameter in the high temperature 
susceptibility for CeCoIn$_5$ and other materials in which crystal field effects 
play out at substantially higher temperatures.

We conclude that once one has identified $T^*$ for a given heavy electron 
material, the interacting components necessary to understand its transport 
behavior at all temperatures as well as the 
appearance of quantum order below $T_0$ can now be specified. There will
in general be three: a heavy electron non-Landau Kondo Fermi liquid 
whose quasiparticles have the temperature dependent effective mass given by
Eq.~(\ref{Eq:KLmass}); a "light" electron Landau Fermi liquid corresponding 
to electrons on those parts of the expanded Fermi surface that have not hybridized 
individually or collectively; and a residual spin liquid made up 
of the screened interacting local moments. So to understand the 
behavior of heavy electron materials between $T^*$ and $T_0$ one needs first 
of all to combine our scaling description of the Kondo liquid behavior and 
the two fluid 
model to follow the temperature evolution of all three components, while it 
is their mutual interaction, in the presence of global hybridization, that 
determines transport properties above $T_0$, and the nature of the ordered 
state below $T_0$.

The work presented in this communication was stimulated in large part by discussions 
at the NSF-supported August 2007 ICAM Workshop on the 115 Materials 
(http://www.i2cam.org/conference/1-1-5 materials) and we wish to thank our colleagues 
there for many stimulating discussions then and subsequently on these and related 
topics. We thank Nick Curro for calling our attention to the proportionality between 
the Knight shift anomaly and tunneling  results, Laura Greene, Kristjan Haule, 
Mike Hundley, Gabi Kotliar, Satoru Nakatsuji, Wan Kyu Park, and Joe Thompson for 
communicating their results to us in advance of publication and for giving us 
permission to report on these. This research was supported by ICAM, in part 
through an ICAM Fellowship (Y.Y.), UC Davis, the Department of Energy, and the 
National Science Foundation.

\end{document}